# An Efficient Method For Multichannel Wireless Mesh Networks With Pulse Coupled Neural Network

S.Sobana
Assistant professor
Department of ECE
PSNA College of Engg & Tech-Dindigul 624622

S.Krishna Prabha
Associate professor
Department of Mathematics
PSNA College of Engg & Tech-Dindigul 624622

*Abstract*—Multi cast communication is a key technology for wireless mesh networks. Multicast provides efficient data distribution among a group of nodes, Generally sensor networks and MANETs uses multicast algorithms which are designed to be energy efficient and to achieve optimal route discovery among mobile nodes whereas wireless mesh networks needs to maximize throughput. Here we propose two multicast algorithms: The Level Channel Assignment (LCA) algorithm and the Multi-Channel Multicast (MCM) algorithm to improve the throughput for multichannel sand multi interface mesh networks. The algorithm builds efficient multicast trees by minimizing the number of relay nodes and total hop count distance of the trees. Shortest path computation is a classical combinatorial optimization problem. Neural networks have been used for processing path optimization problem. Pulse Coupled Neural Networks (PCNNS) suffer from high computational cast for very long paths we propose a new PCNN modal called dual source PCNN (DSPCNN) which can improve the computational efficiency two auto waves are produced by DSPCNN one comes from source neuron and other from goal neuron when the auto waves from these two sources meet the DSPCNN stops and then the shortest path is found by backtracking the two auto waves.

## I. INTRODUCTION

Unlike mobile adhoc networks or wireless sensor networks route recovery are energy efficiency is not the major concern for mesh network due to limited mobility and the rechargeable characteristics of mesh nodes. Moreover supporting major applications such as video on demand poses a significant challenge for the limited bandwidth of WMNs it is necessary to design an effective multicast algorithm for mesh networks. It improves the system throughput by allowing simultaneous close-by transmissions with multichannel and multi – interfaces. It assigns all the available channels to the interfaces instead of just the non-overlapping channels.

We propose level channel assignment algorithm multichannel multicast algorithm to improve throughput for multichannel and multi interface mesh networks. Our design builds a new multicast backbone - tree mesh which partitions mesh network into different levels based on the Breadth First Search (BFS), and then heuristically assigns channel to different interfaces. The Pulse Coupled Neural Network is a very active neural network .The PCNN is modified so that the output pulses decay in times. These modified PCNN models need fewer neurons than other approach. This paper proposes a faster PCNN model, which can improve the computational efficiency significantly.

## II. LEVEL CHANNEL ASSIGNMENT ALGORITHM

The nodes obtain their level information. The BFS is used to traverse the whole network. All the nodes are portioned into different levels according to the hop count distances between the source and the nodes.

If node a (in level i) and b (in level i+1) are within each other's' communication range, then 'a' is called the parent of 'b', and 'b' is called the child of 'a'.

We build a multicast tree based on the node level information. Initially, the source and all the receivers are included in the tree. Then, for each multireceiver v, if one of its parents is a tree node then connect it with that parent, and stop. Otherwise randomly choose one of its parents, say fv, as relay node on the tree, and connect v and fv. Afterwards, we try to find out the relay node for fv recursively. The process repeats until all the multireceivers are included in the multicast tree.

The tree nodes decide their channel assignment with the level information.

- The source node (level 0) only uses one interference, which is assigned channel 0. This interference is responsible for sending packets to the tree nodes in level 1.

- The internal tree node in level i (i≥1) uses two interfaces: one is assigned channel i-1, which is used to receive packets from the upper level; the other is assigned channel 1, which is used to forward the packets to the tree nodes at level i+1.

- The leaf in the level I (i≥1) uses two interfaces: one uses Channel i-1 to receive the packets from level i-1,the other uses channel I to forward the packets to the mesh clients within the communication range that desire to receive the packets.

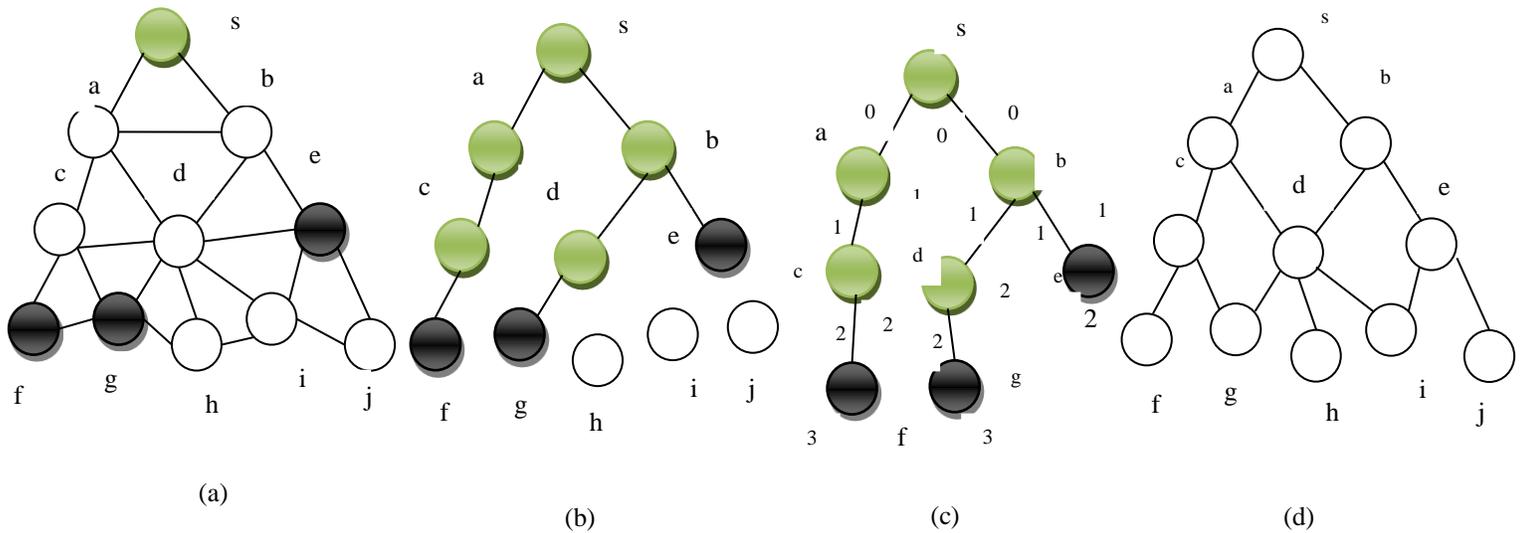

Fig.1.An example for LCA and tree mesh,(a)Network topology ,(b) multicast tree, (c) channel assignment, and (d) tree mesh

For example in fig.1, the node s is the source and nodes f, g, e are the multireceivers. In fig1.a {s, f, g, and e} are included in the multicast tree. Since nodes of g's parents are tree nodes, it randomly selects d as a parent node and connects node g with d. Then choose d's parent b as a tree node and connect d with b. Since b's parent s is a tree node connect b with s. Next, we start from multireceiver e. Connect e with its parent node b and stop because b is already connected with tree node s. Similarly the third multireceiver f, connect f with c, c with a and then a with s. Thus the tree construction is completed by connecting all the receivers with the tree.

The LCA algorithm has two advantages: simple implementation and throughput improvement. At the same time the use of multiple channels reduces the close-by interference and allows more simultaneous transmissions.

To improve the system throughput by the following ways, first LCA cannot diminish the reference among the same level s since it uses the same channel at the same level. Second, when the number of available channels is more than that of the levels, some channels will not be utilized, which is a waste of channel diversity. Third, the channel assignment does not take the overlap property of the two adjacent channels into account. For all I, channel and channel i+1 are adjacent in frequency, so they partially interfere with each other. Thus, the channel I for level i sti ll has some interference effect with the channel i+1 for level i +1.

### III. MULTICHANNEL MULTICAST

#### A. Algorithm

To improve the system throughput, the MCM algorithm is proposed to minimize the number of relay nodes and the hop count distances between the source and the destinations, and further reduce the interference by exploiting all the partially overlapping channels instead of just the orthogonal channels.

#### B. Construction of multicast protocol

When all the Nodes are multireceivers, the multicast problem becomes the broadcast problem. We can say that the broadcast is a special case of multicast. The broadcast structure in the mesh network is built by the following steps.

After the BFS traversal, all the nodes are divided into different levels. Delete the edges between any two nodes of the same level, with which we get the elementary communication structure "tree mesh".Fig.1a and 1d given an example of the original network topology and its corresponding tree mesh.

Identify the minimal number of relay nodes that form the broadcast tree. Using more relay node means more transmissions in the network. Because the number of available channel is limited by current technical conditions, more transmissions would result in more interference and result in more bandwidth cost. Hence, minimizing the multicast tree size helps to improve the throughput. The purpose of this step is to identify the relay node for a node that has more than one parent nodes so that the number of relay node is minimal.

#### C. Structure of Multicast protocol

In broadcast structure unnecessary branches are present if the destinations do not involve all the nodes. Hence, we propose to construct a slim structure y using the MCM Tree Construction algorithm.

The goal of the algorithm is to discover the minimal number of relay nodes needed to construct a multicast tree. The search process starts from the bottom to the top. We use a simple example to explain the process in a tree mesh in Fig.3a, where nodes 6, 7, and 8 are the multireceivers. First select node 4 at level 2 because it covers all the multireceivers at level 3.Next select node 2 at level 1, which covers all the multi receivers and the relay node at level 2.By doing these steps finally we get the multicast tree in Fig.3b

#### D. Channel Assignment

Multi receivers can be connected with the gateway through mini al hop count distance as discussed earlier. Now we discuss about how to assign channels to the interference of tree nodes, for that we propose two allocation algorithms: ascending channel allocation and heuristic channel assignment.

*a) Ascending channel Allocation*

The interference that a node uses to receive packets from its relay node at the upper layer, called as Receiver Interference (RI), is disjoint from the interference the node uses to forward packets to children, called Send-Interference (SI). To guarantee that the relay node can communicate with its children, each node's RI is associated with the SI of its relay node, i.e., they should be assigned the same channel.

The algorithm is explained as follows: From the top to bottom in the tree, the channels are assigned to the interfaces in the ascending order until the maximum channel number is reached, then start from the channel 0 again. Although simple, this approach avoids the situation that the same channel is assigned to two nearby links that interfere with each other. In Fig .4, the numbers of orthogonal channels are three, the number above the node represents the channel number used for its RI, and the number below the node represents the channel number for its SI.

*b) Heuristic Channel Assignment*

We noticed that the interference range decreases with the increase of the channel separation for two wireless links which have short physical distance. To minimize the sum of the interference area of all the transmissions this algorithm is proposed.

We use IR (uv) to indicate the interference range of sender u of one link with respect to sender v of another link According to our consideration all the have the same transmission range R, IR (u v) =R* δ [iu-iv]. iu and iv are the channels of u and v for their SIs, and δ t is the interference factor. When allocating a channel for relay node u, the channel assignment should take a channel that minimize the sum of the square of the IRs between u and u's neighboring relay nodes, i.e., minimize

$$\sum_{v \in N(u)} IR^2(u_v)$$

,where N(u) is the set of neighboring relay nodes of u. This is because the bigger the interference area means bigger chance two transmissions may interfere. The interference area is approximated as a circle whose area is determined by IR2 (uv). Since

$$\sum_{v \in N(u)} IR^2(u_v) = \sum_{v \in N(u)} \left( R * \delta_{|i_u - i_v|} \right)^2 ,$$

The heuristic Channel assignment is used to minimize

$$\sum_{v \in N(u)} \left( R * \delta_{|i_u - i_v|} \right)^2$$

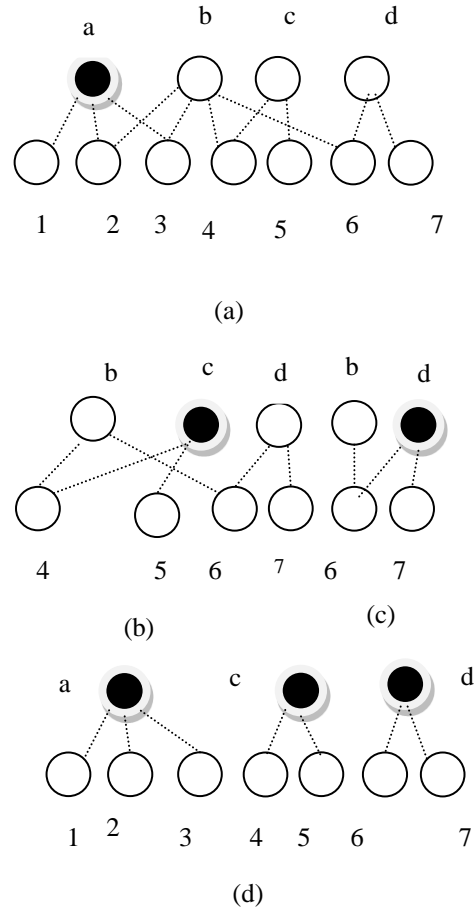

Figure 2. Relay node search example

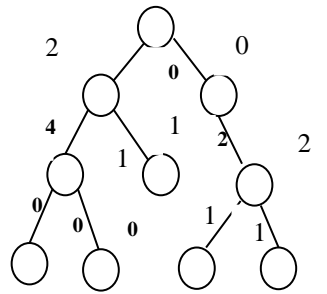

Fig.4. Ascending channel allocation example

## IV. DSPCNN MODEL

*A. Preliminaries*

The input to the preprocessing stage is an undirected graph G = (V, E) with n vertices and m edges, and non-negative lengths l(e) for each edge e. Another two inputs are a source node s and a goal node g. The goal of this algorithm is to find a shortest path from s to g. Let dist(s, g) denote the shortest path length from s to g with respect to l. i.e. dist(s, g) =dist (g, s).

*B. Model Of DSPCNN*

To compute the point-point shortest path more efficiently a Dual Source Pulse Coupled Neural Network (DSPCNN) model is proposed. This model can produce two Auto waves from two different firing sources. At t=t0, the source neuron and goal neuron fire and emit pulses simultaneously. Then, the two Auto waves propagate in parallel by their neighboring neurons at next instant till they meet together. In order to differentiate two auto waves, the auto wave propagating from source neuron is denoted as Ps, and the auto wave propagating from goal is denoted as Pg. If neuron fires on the simulating of Ps auto wave it outputs Ps pulses. If a neuron fires on the simulating of Pg auto wave, it output Pg pulses. If a neuron fires on the simulating of both Ps and Pg pulses, it indicates that the two auto waves meet and the model should stop.

*C. Shortest Path Computation Using DSPCNN*

To compute the shortest path for networks, first they are all transformed to a graph for further processing. The next step is to map the graph into DSPCNN model. Each node in the graph corresponds to a DSPCNN neuron, and each edge associated with a link between neurons. The cost of an edge can be viewed as an external input for the two neurons connected by the edge.

During time t=0, Source neuron and Goal neuron fire simultaneously. Then the auto waves Ps and Pg from the firing sources propagate to their neighbors. A variant meeting is used to determine whether the two auto waves meet together, and the meeting neutron is denoted by Nm. If Nj fires on the simulation of Ni, we call Ni is the precursor of neuron Nj.

## V. CONCLUSION

In our paper we investigate the multicast algorithm wireless networks. In order to achieve efficient multicast in WMNs, two multicast algorithms are proposed by using multichannel and multi interfaces.

These algorithms are focused on increasing the throughput and decreasing delay. With neural networks the proposed DSPCNN is used to achieve higher efficiency and involve lower search space, which can save the run time significantly.

ACKNOWLEDGMENTS

We would like to thank the anonymous reviewers for their helpful comments. We would like to thank our mother for her kind support. We would like to thank our institution and management for their cooperation.

AUTHORS PROFILE

S.SOBANA.-received the B.E (with distinction) in Electronics andCommunication Engineering and M.E (with distinction) in Applied Electronics from RV.S College of Engineering and Technology, Anna University Chennai, Tamil Nadu, India .in 2005 and 2007 respectively. Currently working as an assistant professor in Electronics and Communication Engineering Department, P.S.N.A College of Engineering and Technology, Tamil Nadu, India. Her research mainly focuses on ad hoc networks, congestion control, power management techniques in wireless networks.

S.Krishna Prabha-receiver her B.Sc Mathematics degree in 2000 from G.T.N Arts College, MK University, Tamil Nadu, India. MSc and MPhil degree in Mathematics from M.K. University, Tamil Nadu, India in 2002 and 2004 respectively. Currently she is pursuing the M.E degree from System Engineering and Operation Research department, Anna University, Trichy, Tamil Nadu, India. Currently working as an associate professor in Department of Mathematics, P.S.N.A College of Engineering and Technology, Tamil Nadu, India. Her research interests include graph theory, operation research, boundary value problems, ad hoc networks.